Article

# A method to correct the temporal drift of single photon detectors, based on asynchronous quantum ghost imaging

Carsten Pitsch [1,*], Dominik Walter [1], Leonardo Gasparini [2], Helge Bürsing [1], Marc Eichhorn [1,3]

1. Fraunhofer Institute of Optronics, System Technologies and Image Exploitation (IOSB), Gutleuthausstr. 1, 76275 Ettlingen, Germany
2. Fondazione Bruno Kessler, Center for Sensors & Devices, Integrated Readout ASICs & Image Sensors, Via Sommarive 18-Povo, 38123 Trento (TN), Italy
3. Institute of Control Systems (IRS), Karlsruhe Institute of Technology, Fritz-Haber-Weg 1, 76131 Karlsruhe, Germany
* Correspondence: carsten.pitsch@iosb.fraunhofer.de

**Abstract:** Single photon detection and timing gathered increasing interest in the last few years due to both its necessity in the field of quantum sensing and the advantages of single quanta detection in the field of low level light imaging.

While simple bucket detectors are mature enough for commercial applications, more complex imaging detectors are still a field of research with mostly prototype level detectors. A major problem in these detectors is the implementation of in-pixel timing circuitry, especially for two-dimensional imagers.

One of the most promising approaches is the use of voltage controlled ring resonators in every pixel. Each of those is running independently, based on a voltage supplied by a global reference. However, this yields the problem that across the chip the supply voltage can change, which in turn changes the period of the ring resonator. Due to additional parasitic effects, this problem can worsen with increasing measurement time, leading to a drift of the timing information.

We present here a method to identify and correct such temporal drifts of single photon detectors, based on asynchronous quantum ghost imaging. We also show the effect of this correction on a recent QGI measurement from our group.

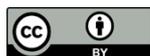

## 1. Introduction

Single photon detectors have advanced significantly in recent years, due to their need in both low-level-light and quantum applications. Current single photon avalanche diode (SPAD) cameras, based on photon counting, already achieve resolutions in the megapixel regime. However, for many applications the temporal information of the photons is crucial, such as LIDAR or quantum sensing based on coincidence sensing.

For these detectors dedicated time-to-digital converters (TDC) are needed, which can be realized in a variety of designs [1]. While each of these designs has its own dedicated issues, keeping the timing information consistent is a challenge for each. In the case of 2D detectors this proves especially challenging, since here independent TDCs are usually used rather than global, clock based TDCs. This allows to reduce TDC size, therefore increasing fill factor, but leads to TDC dependent nonlinearities, which can prove to be a major limitation for a variety of applications. These nonlinearities usually lead to either jitter in the recording or falsified information, i.e. in ToF-LIDAR applications.

In this work we analyze and correct the temporal nonlinearities of a 2D SPAD array with TDCs based on ring resonators. These detectors have individual TDCs for each pixel/

group of pixels, allowing ps temporal resolution while maintaining a high fill factor. This leads to a pixel-dependent correction of the timing information. However, the correction routine is not limited by this, but by the deterministic nature of the influence/ parameter under investigation.

**2. Sources of the temporal drift**

Multiple architectures exist for in-pixel timestamping circuits in a time-resolved, single-photon image sensor, including analog [2] and digital implementations. The latter ones are typically based on ring oscillators and in this case they can be divided into two main groups: one based on a global ring whose phases are distributed across the entire array [3] and another one in which each pixel has its own replica of a ring oscillator [4]. With a local ring, power consumption is optimized thanks to the reversed start-stop operation as only pixels that detect a photon (which are typically a small fraction of the entire array) draw current from the supply, while area compactness is achieved by using a small technology node [5], or sacrificing some resolution [6, 7].

In theory, photons detected at the same time in different pixels of the detector should return the same timestamp. In reality, the timing information from different pixels will exhibit (usually small) variations from non-idealities, resulting in (slightly) different timing information for every pixel/TDC. Many effects can influence this mismatch, such as jitter, the large power consumption of the TDCs, mismatches among the timestamping channels and skew in the distribution of the common reference signal(s). While jitter is a random process, which can only be analyzed statistically, the influence of the other parameters can be deterministic and therefore be addressed in the detector design or its post-processing.

The power distribution network affects the uniformity of the oscillators due to the voltage drop that occurs when the ring oscillators of the TDC are running. This drop is caused by the current I which flows in the metal connections of the TDC, having a non-zero resistance R. Thus it is also referred to as "IR drop". Figure 1 shows a pictorial representation of the power distribution network of a SPAD imager and the associated IR drop. Typically, a ring of power supply (VDD) and ground (GND) pairs surrounds the pixel array, which are then distributed horizontally and/or vertically to all pixels avoiding routing them over the active area.

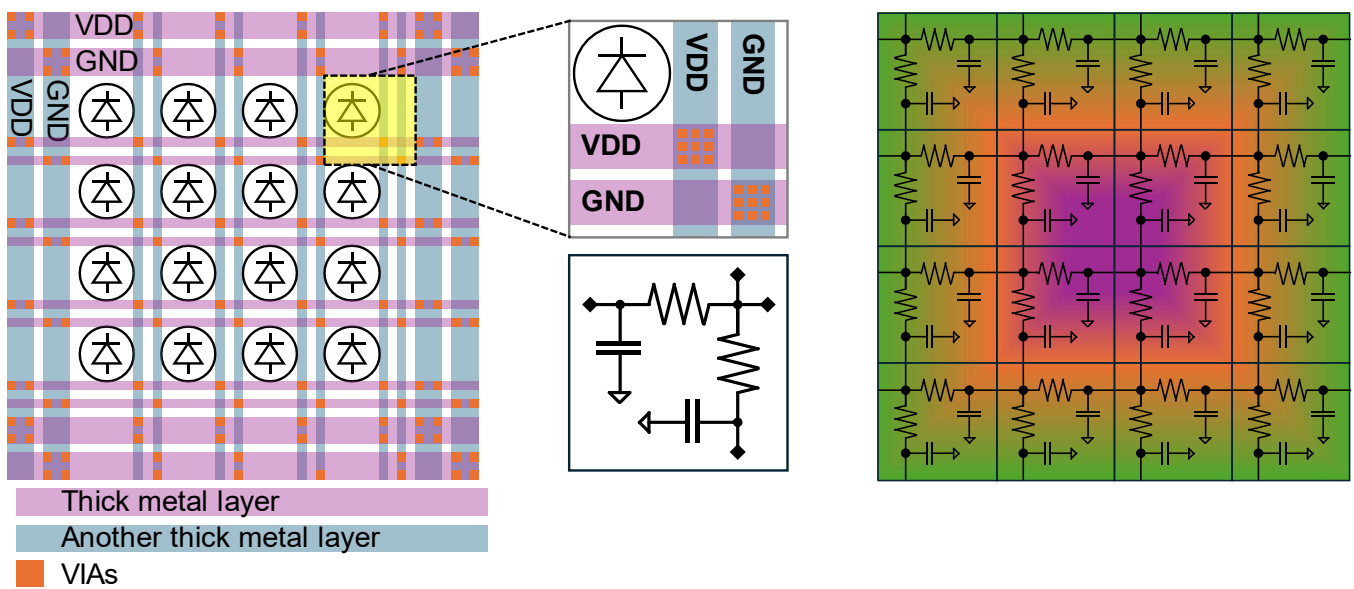

**Figure 1.** Pictorial representation of the power distribution network (left), with pixel zoom-in (center, top), an RC model of a pixel (center, bottom). The resulting network leads to a non-uniform IR drop which is maximum at the center of the array.

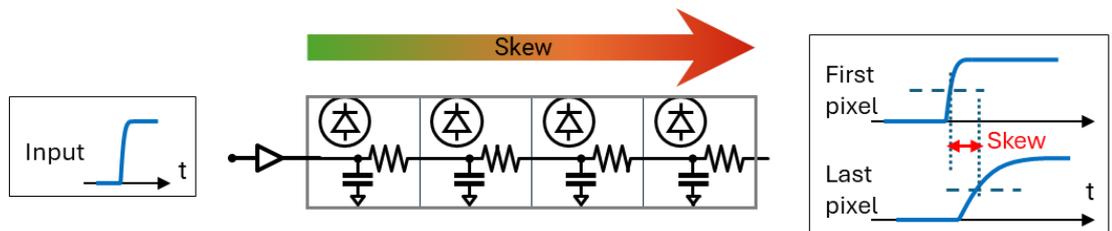

**Figure 2.** Concept of skew caused by the propagation of the timing signal from the periphery of the array to the pixels.

Each pixel contributes to the resistive and capacitive load according to the employed technology, the metal layers being used, and the design layout. The consequence of such a drop is a reduced supply voltage and/or an increased ground level, and the intensity of the effect is inversely proportional to the distance from the center of the array. This is mitigated by the presence of a regulating transistor for the VDD, and gets worse when a large number of TDCs run in parallel, thus posing a limit on the maximum number of detectable photons per observation window. To limit the drop, a small resistance and a large capacitance are desirable, but the need for a high fill factor and a small pixel pitch (required for scalability to large arrays) limit the width of the metal connections.

The manufacturing process is then causing small differences between the transistor layout and the fabricated counterpart, which introduce ring-to-ring mismatches in the frequency of oscillation and non-linearities in the TDC transfer function. These can be mitigated in the design phase, i.e using non-minimum transistor size, and calibrated in post-processing, since the behavior of a specific TDC is determined during the fabrication phase and remains unchanged during operation.

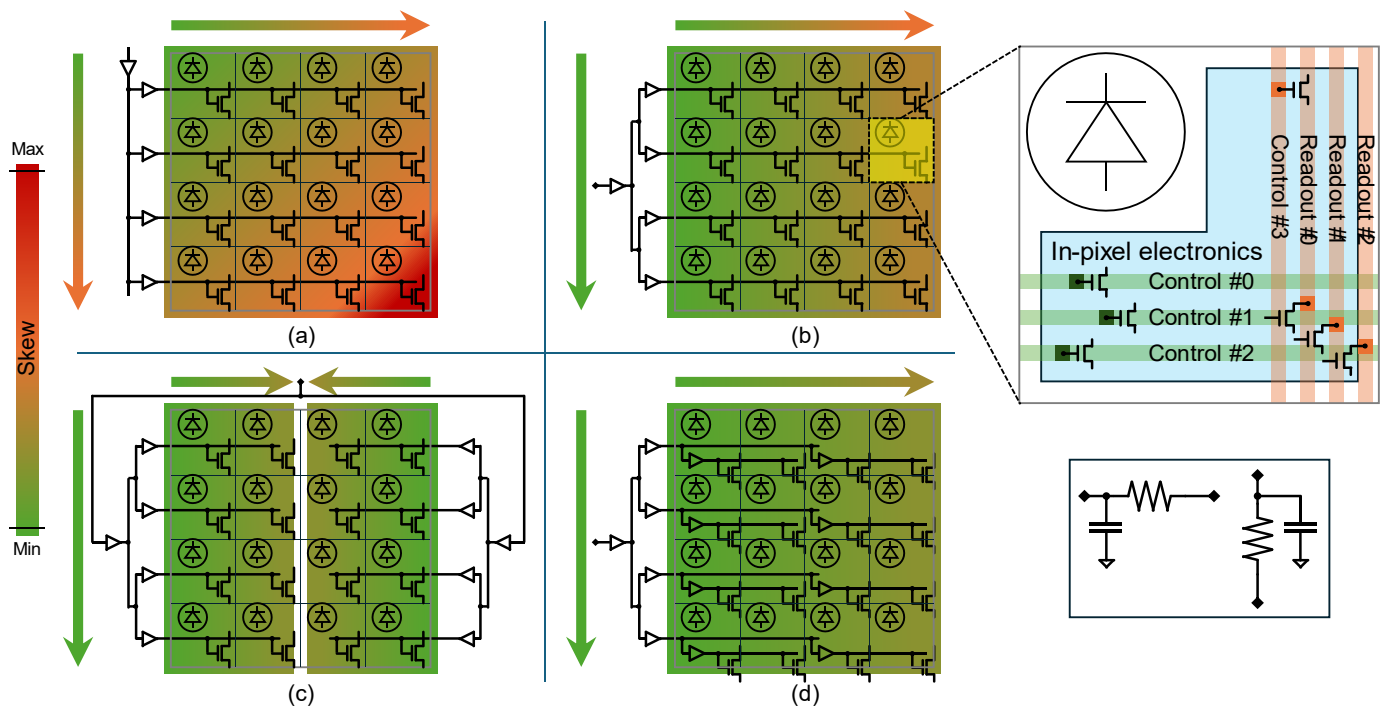

**Figure 3.** Pictorial representation of the distribution of control signals across the array and associated skew across the array. The worst-case skew depends on the architecture of the driving logic. Here, four common architectures are shown: (a) basic row driver, (b) equalized, row-level buffer tree, (c) left & right driver, (d) array-level buffer tree. On the right a pixel zoom-in (top) is shown highlighting the load of control and readout nets, along with the associated RC model (bottom).

Another source of non-ideality is the distribution of the reference timing signal which, despite being a global signal, is routed from the periphery to the array, leading to a delay between the pixels. The variation, known as *skew*, can be analyzed by modeling the distribution of the control signal as a transmission line, consisting of the resistive and capacitive load of each pixel. The resistance in this model is caused by the metal connections itself, while the capacitance includes the parasitic effects of the gates and the metal-to-metal/metal-to-substrate connections. These parasitic effects lead to a variation in the control signal, smearing its edges towards later pixels. As the receiver logic is based on edge detection, this leads to a shift in the detection of the control signal as depicted in Figure 2.

To mitigate this, the overall resistance-capacitance (RC) needs to be minimized. While the capacitance C is typically dominated by the transistor gates, and thus cannot be reduced significantly, the resistance R is determined by the width of the metal connections. However, this width also impacts detector performance, as there have to be a lot of signals in a small area. This leads to an optimum width of these connections.

The skew can further be reduced by properly designing the architecture of the driving logic, as shown in Figure 3. Even for small array sizes, a single buffer can't provide enough driving power. Thus the most simple configuration includes a two-level buffer configuration, in which a main buffer drives a set of row-level buffers. Here, the skew shows a vertical gradient due to the distribution of the signal from the main buffer to the row buffers, and a horizontal gradient due to the row buffers driving the pixels. The vertical gradient can be strongly mitigated by using a buffer tree to distribute the signal to the row-buffers, at the cost of a higher complexity and an increased power consumption. If the horizontal skew is still too large, the driver can be duplicated on the opposite side of the array, so that each driver serves half of the pixels. Alternatively, the tree configuration can be extended inside the pixel array, so that the row-buffers drive a sub-set of the pixels in each row, and the pixels then regenerate the signal and redistribute it locally.

## 3 Correction approach

In order to measure the temporal drift of the detector under evaluation we exploit the temporal correlation of photon pairs created by spontaneous parametric downconversion (SPDC). This correlation is usually used in quantum applications, like quantum ghost imaging [8, 9] in order to identify the pairs by coincidence detection.

In our approach we use this correlation to obtain a timing reference for the detections of the SPAD detector by the appropriate idler detections. By using the asynchronous detection approach described in [10, 11], we are able to store all detections of the measurement digitally, along with all relevant parameters of the detector. This enables us to analyze all dependencies of the coincidence detection on these parameters and with that their influence on the timing of the detection.

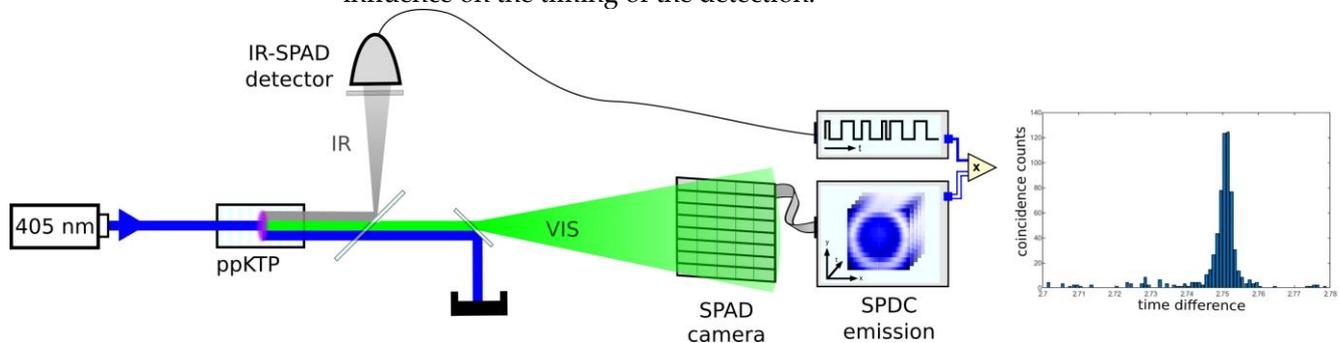

**Figure 4.** The measurement setup to determine the pixel-dependent temporal drift of a SPAD imager [6]. The setup is based on previously published setups for asynchronous quantum ghost imaging [10, 11]. The emission spot on the camera should illuminate the complete detector aperture in order to perform coincidence analysis over the whole chip. This should also be verified for the idler detector, especially in case of fiber coupling.

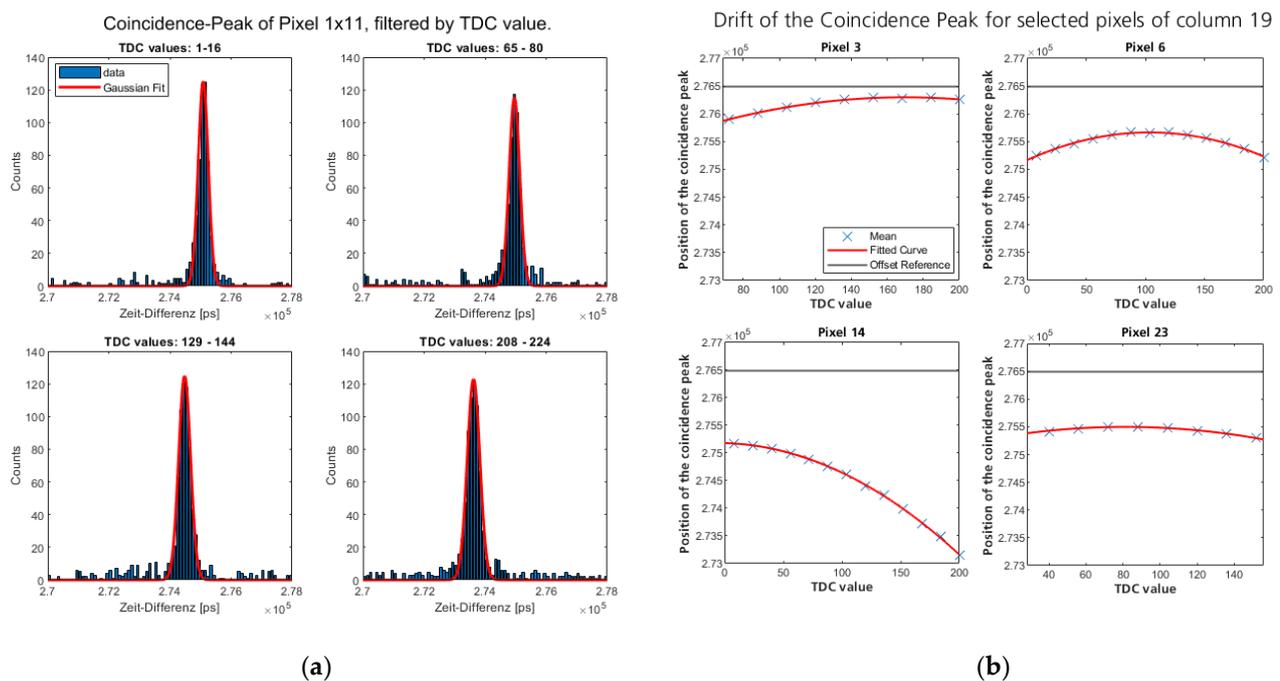

**Figure 5.** Drift of the coincidence peak over measurement time for selected pixels. (a) Coincidence evaluation for groups of TDC values of a single pixel (Column 1, Row 11). 16 TDC values were grouped together to ensure enough data for a good fit function. The dependence of the peak position on the TDC value is clearly visible and results in broadening of the peak. (b) Evolution of the mean of the Gaussian fit over the TDC value for selected pixels. The drift was estimated by a polynomial function of second degree. To avoid inaccuracies with the fit, only TDC values with enough data were evaluated. Because of this, pixel 23 is only evaluated until TDC value 160. Also shown is the global reference value used to create the lookup table for correction.

### 3.1. Measurement system

The measurement system is an adapted version of the QGI system described in [11], the data of which is also used in chapter 4 to show the improvement of the system.

As photon-pair-source a 2 mm long periodically poled KTP crystal is used, which is pumped by a 405 nm laser. The crystals poling period of 4.25µm allows collinear phase matching for non-degenerate photon pairs at 550 nm (signal) and 1550 nm (idler), allowing the use of readily available, low-jitter infrared SPADs as timing reference (ID230, [12]). For synchronization a high-resolution single photon counting board (TimeHarp 260 PICO, [13]) is connected to both detectors to enable timestamping every detection with ps resolution.

### 3.2. Drift Determination

By comparing the timestamps of each detection, coincidence analysis is enabled. This analysis can be filtered by each parameter individually or specific parameter combinations. In particular it allows to filter the detections by both pixel and TDC value. This allows to analyze both the difference in base resolution of each pixels TDC and its behavior over measurement time. In case more information is provided (e.g. START/STOP codes for fine resolution as in [14]), the influence of these parameters can also be analyzed.

As the uncertainty in the correlation of the photons and the timing jitter of the reference detectors are smaller than the resolution of the detector under investigation (~210 ps resolution), the coincidence evaluation allows to determine any influence of the available parameters on the temporal behavior of the detection. It allows this analysis not just in terms of drift of the timing information, but e.g. also in timing jitter.

## 4 Results

In order to test the approach and validate the improvement, we performed the correction on recently published measurements, detailed in [11].

To do so, we built the measurement setup shown in Figure 4, operating with a 20 mW pulsed laser as pump and testing the detector used in [11]. In order to obtain a high precision, we performed continuous measurements over 4 days. Due to limitations in the processible amount of data, individual measurements were limited to 2 hours and then automatically restarted. This also ensures repeatability of the drift and correction over different measurements.

For further processing each of the measurements was evaluated individually, while being filtered by pixel. This isolates the individual TDCs of the sensor, which are responsible for the timing drift. The temporal behavior of the TDCs is then analyzed by filtering the coincidence data again, this time by the value of the TDC itself. In order to reduce the data to evaluate the complete set of measurements, the coincidence window is split into 100 ps large sections, while for every section the amount of coincidence counts is saved, This allows to simply add up the results of each measurement, increasing the amount of data and thus the validity of the correction.

The evolution of the coincidence peak over TDC value is exemplary shown in Figure 5(a) for a single pixel. With the data shown, the drift was analyzed by fitting a normal Gaussian distribution to the peak. To ensure enough data yield for a valid fit, the TDCs were grouped at 16 groups of 16 TDC values each.

The mean of this fit gives the temporal drift of the coincidence and is exemplary shown for selected pixels of a column of the detector in Figure 5(b). Due to the detector architecture (see [6]), the highest TDC values (~ above 200) are unlikely to occur, resulting in too few data for a valid fit. Thus, datasets of the TDC groups were filtered by the amount of data before being considered for the amount of temporal drift. The temporal drift was then fitted with a polynomial function of second order.

In a first approach for correction, the parameters of the fit were passed to the evaluation code of the original dataset of [11] and the offset of each detection was calculated. However, this highly increased the runtime of the code and thus wasn't fit for usage. Instead, we calculated the offset of the drift to a global reference value for every possible TDC value and created a lookup table with this. For every detection we then add the dedicated lookup value (determined by pixel number and TDC value) to our detection time, thus eliminating the drift.

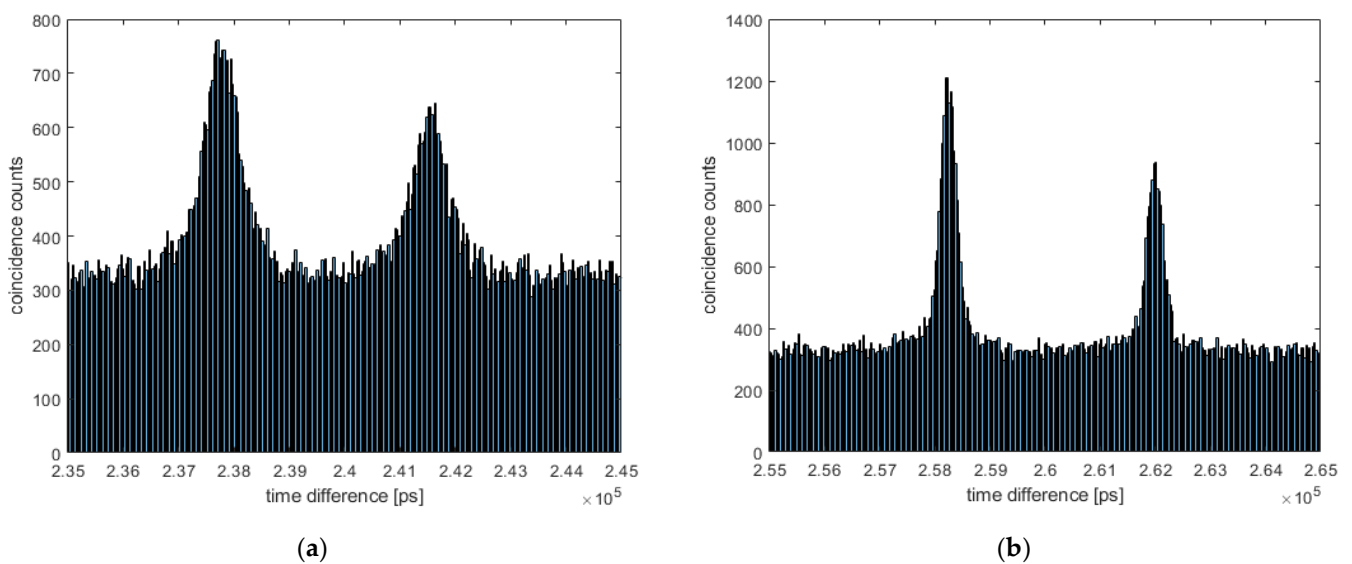

(a)  (b)

**Figure 6.** Coincidence evaluation with different corrections. (a) Coincidence peak with linear correction of the TDC values, as published in [11]. The peak is about 2 ns wide with an FWHM of about 700 ps. (b) Coincidence peak with the proposed drift correction, yielding a full peak width of about 700 ps and a FWHM of about 300 ps.

Applying this correction, the coincidence peak could be improved from a full width of about 2 ns with a linear correction (as performed in [11]) to about 700 ps, as shown in Figure 6. This peak width already comes close to the resolution of the detectors TDCs of ~210 ps [6]. We presume this peak width is currently also limited by the resolution of the reference detector in the idler, which shows a jitter of ~150 ps [12], and the TDC resolution used for the correction, currently at 16 values. It should be mentioned that the linear correction in [11] was already refined using coincidence data. The drift correction shown here however, was based on a coarse linear correction, detailed in [6]. With this coarse correction a coincidence peak of about 3 ns was measured

## 5. Outlook

We have shown here a method to accurately determine the temporal behavior of single photon cameras, exploiting the temporal correlation of photon pairs created by SPDC.

This allowed us to increase the temporal resolution of our asynchronous QGI setup by a factor of ~3. The achieved resolution of ~700 ps is already close to the base resolution of the detector under investigation (~210 ps) and the reference detector used (~150 ps). It might be improved further, especially by increasing the resolution of the drift determination by a more detailed analysis of the TDC dependence. Optimally by analyzing each individual TDC value instead of groups of 16 values. For this analysis to be valid however, more data should be acquired.

Another improvement would be the use of a more precise bucket detector, i.e. a super-conducting nanowire single photon detector (SNSPD). Using a silicon detector as bucket might also be an option, but for most SPAD imagers the wavelength of the signal photons should be kept in the 550 nm regime due to detection efficiency of the pixels. The realization of SPDC with this signal wavelength and an idler wavelength in the silicon detection range does lead to problems regarding the source. Due to energy conservation a UV laser would have to be used, leading to problems regarding degradation of the non-linear crystal.

However, the system is not limited by the use of SPDC, but only on the fundamental temporal correlation of the photons. Thus, any source of time-correlated photon states (i.e. also Four-Wave-Mixing) could be used. The correlation must only be verified to be more exact than the resolution of the detectors.

It might also be an option to use lasers with temporally shifted triggers as a method to determine the temporal drift of these detectors. For this a very fast switching laser would be required as well as special filters, due to the sensitivity of the detector. It is therefore questionable whether the timing of such a laser emission scheme is sufficiently precise to achieve the detectors resolution. It should further be noted that such a flood illumination can change the timing properties of a detector, leading to a falsified correction for low-level light applications. Our approach however works with illumination levels similar to other quantum and low-level-light applications.